# Effective Dielectric Response of Dense Ferroelectric Nanocomposites

Oleksandr Pylypchuk
*Institute of Physics, NAS of Ukraine*
46, pr. Nauky, 03028 Kyiv, Ukraine

Serhii Ivanchenko
*Institute for Problems of Materials Science, NAS of Ukraine*
Krjijanovskogo 3, 03142 Kyiv, Ukraine

Yuriy Zagorodniy
*Institute for Problems of Materials Science, NAS of Ukraine*
Krjijanovskogo 3, 03142 Kyiv, Ukraine

Oksana Leschenko
*Institute for Problems of Materials Science, NAS of Ukraine*
Krjijanovskogo 3, 03142 Kyiv, Ukraine

Oleksii Bereznykov
*Institute of Physics, NAS of Ukraine*
46, pr. Nauky, 03028 Kyiv, Ukraine

Mykola Yelisieiev
*Institute of Semiconductor Physics, NAS of Ukraine*
45, pr. Nauky, 03028 Kyiv, Ukraine
mykola.eliseev@gmail.com

Vladimir Poroshin
*Institute of Physics, NAS of Ukraine*
46, pr. Nauky, 03028 Kyiv, Ukraine

*Abstract.* We measured the temperature dependences of the capacity and dielectric losses of the tape-casted dense ferroelectric composites consisting of 28 vol.% BaTiO$_3$ nanoparticles (average size 24 nm) in the polyvinyl butyral polymer and 35 vol.% BaTiO$_3$ microparticles (0.5-0.7 µm) dispersed in the ethyl-cellulose, as well as the ceramics made from the nanoparticles using the annealing at 1250°C. The composite films were covered with Ag electrodes and the measurements were performed in a wide frequency range. Using the Lichtenecker effective media approximation, we analysed the effective dielectric response of the dense ferroelectric composites. To establish the phase state of the nanoparticles, we analysed the static $^{137}$Ba NMR spectra of BaTiO$_3$ nanoparticles in a wide temperature range from 200 to 400 K. Obtained results revealed that the ferroelectric properties of the nanoparticles contribute to the effective dielectric response of the composites in a significantly different way than predicted by the effective media model. This counterintuitive conclusion indicates on the strong crosstalk effects in the dense ferroelectric nanocomposites, which can be very promising for their applications in nanoelectronics and energy storage.

## I. Introduction

Polar-active nanocomposites and colloids, containing ferroelectric nanoparticles of various shapes and sizes, are unique model objects for fundamental research of surface, dimensional, and crosstalk effects in the nanoparticle ensembles [1]. At the same time, the nanocomposites are promising materials for information storage [2,3] and energy harvesting [4], as well as for electrocaloric applications [5,6]. Despite traditional and innovative synthesis methods, control of size, shape, and polar properties is well developed for ferroelectric nanoparticles, with some aspects still containing challenges for the preparation technology and unravelling the mystery for the fundamental theory even in the simplest case of quasi-spherical BaTiO$_3$ nanoparticles with a size of (5–50) nm [7,8,9].

In this work, we investigated the impact of nanoparticle sizes on the effective dielectric properties of suspensions and the polymer-ferroelectric composite films prepared using the particles. From the rheological data, an effective viscosity, flow behavior index, normalized thixotropy/rheopexy degree and hydrocluster size was calculated (table 1). The Ostwald-de Waele law was employed in conjunction with the power law model to assess and model fluid viscosity based on the flow curves of the suspensions. The flow behavior index, indicative of shear thinning or thickening, was also calculated using the power law model. The normalized thixotropy/rheopexy degree, signifying the presence of a renewable structure in the fluid, was determined using the method outlined in [10]. The hydrocluster size, representing the average diameter of the suspension's structural elements, was calculated using the Peclet number equation outlined in [11]. This involved comparing the Peclet number of a known-size powder dispersion with that of a tape casting suspension utilizing the same powder and concentration.

Next, we prepared the tape-casted dense ferroelectric composites consisting of 28 vol.% BaTiO$_3$ nanoparticles (average size 24 nm) in the polyvinyl butyral polymer and 35 vol.% BaTiO$_3$ microparticles (0.5-0.7 µm) dispersed in the ethyl-cellulose, as well as the ceramics made from the nanoparticles using the annealing at 1250°C. The composite films were covered with Ag electrodes and the measurements of the temperature dependences of their capacitance and dielectric losses were performed in the frequency range 100 Hz – 100 kHz.

## II. Experimental Details

The analysis of rheological studies allowed us to draw conclusions about the mechanism of structure formation in the investigated polymer-ferroelectric nanocomposites.

TABLE I. Rheological Properties of Suspensions

| Sample name | Effective viscosity ($\Upsilon$=800 1/s), mPa·s | Flow behavior index, n | Normalized thixotropy/ rheopexy degree | Average hydrocluster size, nm |
|---|---|---|---|---|
| TCS-27 | 108,52 | 1,10 | -1,29 | 87 |
| TCS-34 | 214,03 | 0,98 | -2,20 | 1389 |

The TCS-27 suspension with BaTiO$_3$ nanoparticles, which average size is 24 nm and the size distribution function is very narrow in the range from 17 to 30 nm, exhibits characteristics of a low-viscosity liquid with a slight shear thickening (n > 1) and mild rheopexy behavior. This flow profile is typical for suspensions of nanoparticles dispersed in a polymer solution with low molecular mass [12]. The low viscosity arises from the nature of the polymer solution, which comprises a blend of short-chain Butvar with B75 and a high content of DBP plasticizer, that build in polymer chains additionally reducing their length. Based on the calculated average hydrocluster size, it can be inferred that short polymer chains of PVB contract a few BaTiO$_3$ nanoparticles, generating nano-sized hydroclusters, which behave akin to solid spheres in the flow. This inference finds support in the viscosity curve (see Fig. 1a and Fig. 2), where the TCS-27 nanopowder suspension exhibits nearly Newtonian flow, indicating minimal interaction between structural elements. The observed low shear thickening and rheopexy degree are likely attributed to the high specific surface area of nanoparticles, resulting in increased friction during collisions.

TCS-34 comprises sub-micron BaTiO$_3$ particles dispersed in a polymer solution of long-chain ethyl-cellulose molecules. The use of long-chain polymer leads to increased viscosity and a shear-thinning flow characteristic in the liquid. The elevated viscosity, coupled with the shear-thinning behavior (n < 1), indicates the formation of large hydroclusters capable of deformation and stretching in the shear direction [13] resulting in reduced shear resistance and viscosity (see Fig. 1b and Fig. 2). The unexpected rheopectic flow, uncommon for shear-thinning liquids, could be attributed to the loss of orientation of shear-deformed elongated hydroclusters, resulting in increased flow resistance as the shear rate decreases.

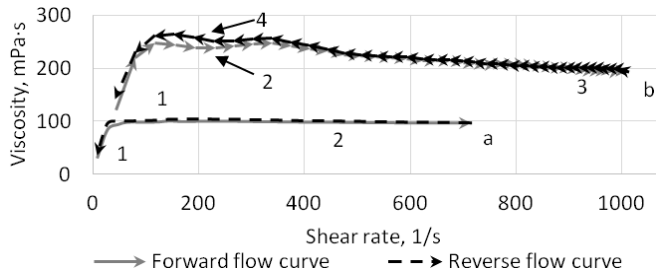

Fig. 1 – Viscosity curves of the nanopowder (a) and micropowder (b) based suspensions.

TCS-27

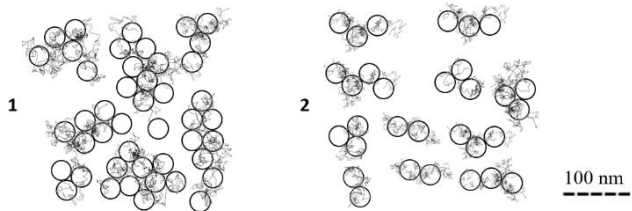

TCS-34

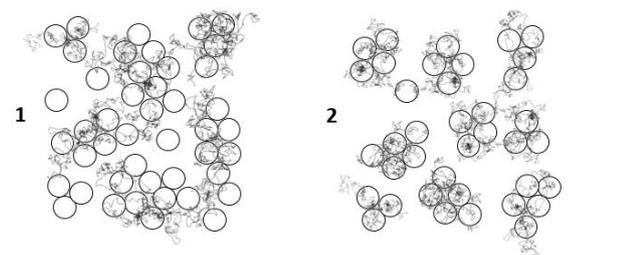

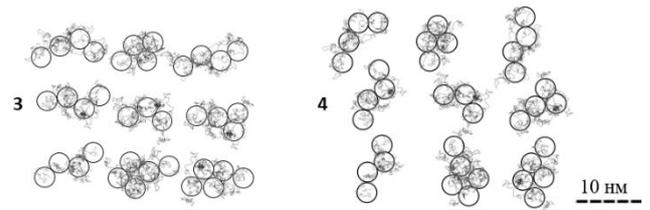

Fig. 2 – Structural states of suspensions with the shear change.

The tape casting of the prepared suspensions was carried out using a tape casting machine with a stationary blade, a carrier speed of 0.55 m/min, and a blade gap of 400 μm for TCS-27 and 200 μm for TCS-34. Films properties represented in the table 2.

TABLE II. PROPERTIES OF THE TAPE-CASTED FILMS

| Sample name | Carrier speed, m/min | Blade gap, nm | Casting shear rate, s$^{-1}$ | Tape thickness, nm | Roughness Rz, nm | Roughness Ra, nm |
|---|---|---|---|---|---|---|
| TCS-27 | 0,55 | 400 | 23 | 18 | 240 | 85 |
| TCS-34 | 0,55 | 200 | 46 | 8 | 590 | 150 |

The reduction in hydrocluster size, achieved through the use of nanopowder, led to a surface roughness parameter twice as low and potentially increased the density of TCS-27 films based on the nanopowder. The reduction in surface roughness suggests improved uniformity, offering potential benefits for applications in multilayer electronics.

To verify this conclusion, we measured the dielectric characteristics of polymer-ferroelectric composites. Also, we studied a pellet made of pure BaTiO$_3$ powder ceramics, annealed at 1250°C. These results we compared with the ones which were obtained for the composites, as well as with the results obtained in other works, for this compound. The composite was in a form of a thin film, plated on top of a metalized textolite plate. Using an LCR-meter, we measured the capacitance (and derived the effective dielectric permittivity from it), as well as the tangent of the dielectric losses, in several samples with different film thickness. These measurements were done for several frequencies, ranging from 100 Hz to 100 kHz, as well as a temperature, from the room point, and up to approximately 180°C.

In order to determine the influence of particle sizes on the properties of studied composites, and considering that the temperature of phase transitions of BaTiO$_3$ nanoparticles can be significantly shifted due to the influence of mechanical stresses and the depolarization field, we probed their phase composition by the $^{137}$Ba NMR. The NMR spectra of the studied samples were recorded on a Bruker Avance II 400-MHz commercial NMR spectrometer in a magnetic field of 9.40 T, corresponding to the Larmor frequency of 44.466 MHz. The $^{137}$Ba NMR spectra were obtained with the conventional 90x−τ−90y−τ spin echo pulse sequence using a four-phase "exorcycle" phase sequence (xx, xy, x–x, x–y) to form echoes with minimal distortions due to antiechoes, ill-refocused signals, and piezoresonances. The p/2-pulse length was typically $t_{\pi/2} = 4\mu$s, the spin-echo delay time $\tau$ was 20 $\mu$s and the repetition delay between scans was 0.15 s. At least 100,000 scans were collected for each temperature to obtain the spectrum of the powdered sample. The spectra were measured at 200, 300, 350 and 400 K. The $^{137}$Ba spectra were referenced against BaF$_2$, which gives a narrow $^{137}$Ba resonance, taken here as the zero for the chemical shift.

### III. EXPERIMENTAL RESULTS

Let us now discuss the dielectric response of the polymer-ferroelectric composites, and then compare them with the response of the ceramics sintered from the same BaTiO$_3$ nanoparticles. As one can see from Fig. 3(a) for TCS-27, the maximum of the dielectric permittivity is present for all curves. At lover frequencies it is around 110°C, while at higher ones it shifts towards 125°C. The maximum value decreases with the increase of the frequency. However, the dielectric losses tangent behaves quite differently with the increase of the temperature. At low frequencies (100 Hz and 120 Hz) a maximum occurs at approximately 55°C, and the minimum is observed slightly before the peak of the permittivity, at 100°C. At the same time, for higher frequencies (1 – 100) kHz, the maximum shifts towards higher temperatures, and the minimum slowly disappears completely. And still, the peak of the losses stays well before the peak of the permittivity, slowly decreasing in its amplitude with the frequency increase.

For TCS-34, the results are shown in Fig. 3(b). The dielectric permittivity has a very sharp maximum at 125°C at low frequencies. At the higher frequencies the maximum stays at the same temperatures and becomes much more diffuse. Its value decreases significantly. The dielectric losses have sharp maxima for 100 Hz and 120 Hz, as well as for 1 kHz, which almost coincide with the maxima of the dielectric permittivity. For higher frequencies, there are no maxima or minima on the losses, and the maximal values of the losses decrease with the frequency increase.

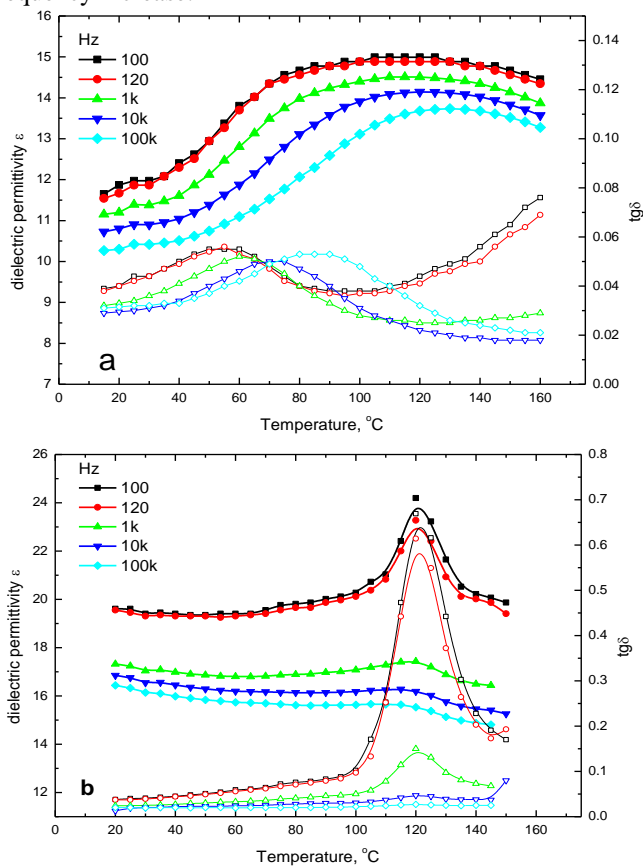

Fig. 3. Temperature dependence of the dielectric permittivity and the losses tangent in the TCS-27 (a) and TCS-34 samples (b) with the Ag electrodes. The permittivity is shown by thick with filled symbols; and the losses are shown by thin curves with empty symbols.

Let us compare the above shown results for dense nanocomposites with the results for a ceramic pellet, which was annealed at 1250°C from the ferroelectric BaTiO$_3$ nanopowder. The dielectric response of the ceramic, shown in Fig. 4, agrees with the well-known response of BaTiO$_3$ ceramics (see e.g., Glass and Lines handbook [14]). However, the maximum on the permittivity, and the subsequent minimum on the losses, occurred at approximately 150°C, which is on (20 – 25)°C bigger than that for a bulk single-crystal. The increase can be related to several reasons, such as domination of tensiled quasi-spherical grains in the pellet and/or some kind of tensiled Vegard strains induced by dopants (e.g., Ca) and/or vacancies. Notably, that compressive strains can lead to the decrease of the transition temperature in quasi-spherical grains.

For an estimate of the effective permittivity, $\varepsilon_{eff}$, of the composite, we can use the Lichtenecker's logarithmic mixture law [15], which is valid for both random inclusions shapes, as well as their orientation. For $\varepsilon_{eff}$, it gives the following equation:

$$log\varepsilon_{eff}^{LR} \approx v_P log\varepsilon_P + (1 - v_P)log\varepsilon_M. \qquad (1)$$

Using Eq. (1), we evaluated the temperature dependence of the $\varepsilon_{eff}$, using the frequency-dependent and temperature-dependent nanoparticle permittivity $\varepsilon_P$ given in the form of the data we obtained for the ceramic pellet, which are shown in Fig.4, and the matrix permittivity $\varepsilon_M = 3$, which is close to the permittivity of polyvinyl butyral (PVB) or ethyl cellulose (ETC) and regarded frequency- and temperature-independent. The fraction of the nanoparticles $v_P$ is (28 – 35)% in accordance with the preparation conditions. We also included the sharp Gaussian-type size distribution function in the calculations of $\varepsilon_P$ with the average size 24 nm and the 3-nm dispersion. The calculated dielectric response is shown in Fig. 5. The size distribution is not significant for the (0.5 - 0.7) μm nanoparticles placed in the ethyl-cellulose. As one can see from the figure, the calculated maximum of the dielectric response is much sharper than the measured maximum of the response shown in Fig. 3(a) and especially 3(b). Moreover, the relative height and the baseline of the maxima in Fig. 5 is significantly higher than the height and baseline shown in Figs. 3(a) and 3(b). Hence, the measured effective dielectric response of the ferroelectric-polymer composites behaves significantly different than predicted by the effective media model. The reason of this discrepancy is not the size distribution of the nanoparticles, but rather their inhomogeneous distribution, as well as interface effects, which can be very important. The interface effects and inhomogeneity can cause the cross-talk effects, which are totally ignored in the linear effective media models.

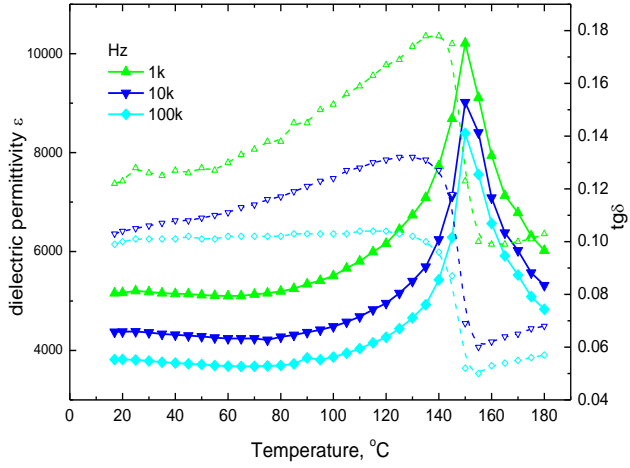

Fig. 4. Temperature dependence of the permittivity (thick curves with filled symbols) and the losses (thin curves with empty symbols) for a BaTiO$_3$ pellet.

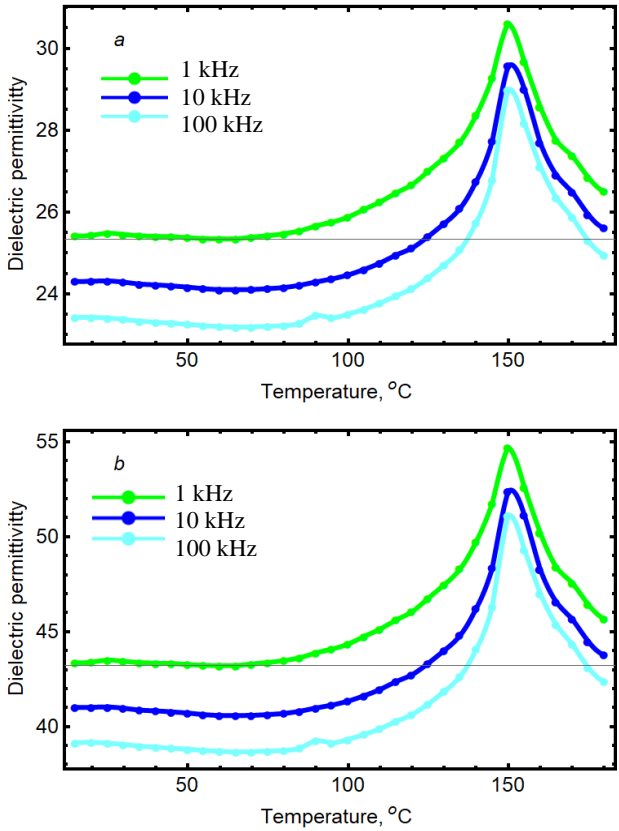

Fig. 5. Temperature dependence of the permittivity for a composite, calculated form Eq.(1) for $v_P = 0.28$ (a) and for $v_P = 0.35$ (b) at different frequencies 1 kHz, 10 kHz and 100 kHz.

The $^{137}$Ba NMR spectra of BaTiO$_3$ nanoparticles are shown in Fig. 6. Due to the ionic character of the chemical bond formed by the barium atom in BaTiO$_3$, the shape of the $^{137}$Ba NMR spectra is largely determined by the interaction of its quadrupole moment (Q = 24.5 fm$^2$) with the electric field gradient (EFG) created by external charges. Only the 1/2 ↔ -1/2 central transition perturbed by the second-order quadrupole interaction of the $^{137}$Ba nuclei with the EFG is observed in Fig. 6. The shift of the central transition due to the solely quadrupole interaction can be expressed as [16]:

$$v_{1/2}^{(2)} = -\frac{v_Q^2}{16 v_L}\left(I(I+1) - \frac{3}{4}\right) f_\eta(\theta, \varphi), \quad (2)$$

where $v_Q = -\frac{3 C_Q}{2I(2I-1)}$, and $C_Q$ is the quadrupole coupling constant whose value $C_Q = e^2 qQ/\hbar$ together with the asymmetry parameter of the EFG: $\eta = \frac{V_{xx} - V_{yy}}{V_{zz}}$ determines the effect of quadrupole interactions on the NMR spectra.

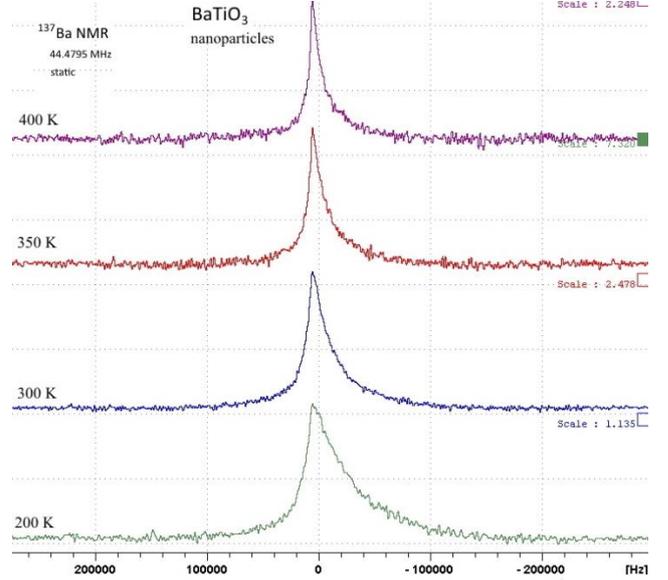

Fig. 6. Static $^{137}$Ba NMR spectra of BaTiO$_3$ nanoparticles obtained at 200, 300, 350 and 400 K.

BaTiO$_3$ ceramics at different temperatures passes through four phases: cubic non-polar phase Pm$\bar{3}$m for temperatures above 393 K, tetragonal P4mm in the temperature range 278-393 K, orthorhombic C2mm between 193-278 K, and rhombohedral R3m below 193 K [17]. According to [18] the asymmetry parameter η is equal to zero in rhombohedral, tetragonal and cubic phases due to the axial symmetry of the EFG, while η is 0.85 in orthorhombic phase, which makes it possible to determine the phase composition based on the shape of the NMR spectra.

The shape of the obtained spectra does not explicitly contain features characteristic of NMR spectra corresponding to different EFG symmetries, but is rather characteristic for the BaTiO$_3$ compound, where the dispersion of the EFG values is present. The line shape analysis of experimentally obtained spectra was carried out under the assumption that the nanoparticles under study have a core-shell structure with a disordered "shell" and "core" with a clearly defined symmetry of the crystal lattice. To model the disordered part of the BaTiO$_3$, the normal distribution of the EFG parameters was used [19]. The best fit to the experimentally obtained spectrum at room temperature is shown in Fig. 7.

As can be seen from Fig. 7, the spectrum can be presented as the sum of three components corresponding to disordered BaTiO$_3$ and BaTiO$_3$ in the orthorhombic and tetragonal phases. The ratio of tetragonal to orthorhombic phase fractions at 300 K is 49:51. At higher temperatures, a narrow line appears, corresponding to cubic symmetry, where the electric field gradients are close to zero. The phases distribution obtained from the NMR data is in good agreement with previously obtained synchrotron x-ray diffractions data for particles 20 and 40 nm in size [20].

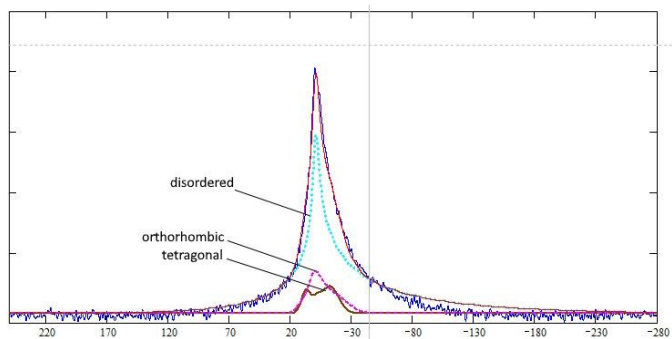

Fig. 7. Fitting of the $^{137}$Ba NMR spectra of BaTiO$_3$ nanoparticles obtained at 300 K by 3 lines corresponding to the disordered BaTiO$_3$ and BaTiO$_3$ in orthorhombic and tetragonal phases.

IV. DISCUSSION

We measured the temperature dependences of the capacity and dielectric losses of the tape-casted dense ferroelectric composites consisting of 28 vol.% BaTiO$_3$ nanoparticles (average size 24 nm) in the polyvinyl butyral polymer and 35 vol.% BaTiO$_3$ microparticles (0.5-0.7 μm) dispersed in the ethyl-cellulose, as well as the ceramics made from the nanoparticles using the annealing at 1250 ºC. The composite films were covered with Ag electrodes and the measurements were performed in a wide frequency range.

Measuring the dielectric losses, as well as the permittivity of these materials, at different frequencies, we were able to determine the following. Two composite samples, with a PVB or an ethyl-cellulose matrixes, were very different in the temperature dependence of the losses and the permittivity.

The first one, with a PVB matrix, has a diffuse maximum on the permittivity, corresponding to a minimum on the losses. But the second one had a sharp peak, with a maximum on the losses at the same temperature, as well. The ceramic pellet shows results, consistent with those, shown in various works or textbooks.

We tried to compare our experimental results for the real part of the dielectric permittivity with an evaluation obtained from the Lichtenecker's mixture law. But, in these results, the peak is still very sharp, and at temperatures slightly higher, then those we saw on the data for TCS-34, for example. Moreover, the measured effective dielectric response of the ferroelectric-polymer composites is significantly different than the predicted by the effective media model. This counterintuitive conclusion indicates on the strong crosstalk effects in the dense ferroelectric nanocomposites, which can be very promising for their applications in nanoelectronics and energy storage.

ACKNOWLEDGMENTS

The theoretical part of the research, materials preparation characterization is sponsored by the NATO Science for Peace and Security Programme under grant SPS G5980 "FRAPCOM" (O.P., Y.Z. and S.I.). Measurements are sponsored in part by the Target Program of the National Academy of Sciences of Ukraine, Project No. 4.8/23-p (O.P., V.P., and V.V.).